\documentclass[%
reprint,%
 amssymb, amsmath,%
 aip,apl,%
]{revtex4-1}

\usepackage{docs}%
\usepackage{bm}%
\usepackage[colorlinks=true,linkcolor=blue, citecolor=red]{hyperref}%
\expandafter\ifx\csname package@font\endcsname\relax\else
 \expandafter\expandafter
 \expandafter\usepackage
 \expandafter\expandafter
 \expandafter{\csname package@font\endcsname}%
\fi
\usepackage{graphicx}

\begin{document}
\title{ Atomistic simulations on ductile-brittle transition in  $<$111$>$ BCC Fe nanowires}

\author{G. Sainath}%

\email{sg@igcar.gov.in}
\affiliation{Deformation and Damage Modelling Section, Materials Development and Technology Division, 
Indira Gandhi Center for Atomic Research, HBNI, Kalpakkam, Tamilnadu -603102, India}%

\author{B.K. Choudhary}

\email{bkc@igcar.gov.in}

\affiliation{Deformation and Damage Modelling Section, Materials Development and Technology Division, 
Indira Gandhi Center for Atomic Research, HBNI, Kalpakkam, Tamilnadu -603102, India}
%\date{}%
%\revised{}%

%\doublespacing
%\begin{onecolabstract}
\begin{abstract}

Molecular dynamics simulations have been performed to understand the influence of temperature on the tensile deformation 
and fracture behavior of $<$111$>$ BCC Fe nanowires. The simulations have been carried out at different temperatures in 
the range 10-1000 K employing a constant strain rate of $1\times$ $10^8$ $s^{-1}$. The results indicate that at low 
temperatures (10-375 K), the nanowires yield through the nucleation of a sharp crack and fails in brittle manner. On the 
other hand, nucleation of multiple 1/2$<$111$>$ dislocations at yielding followed by significant plastic deformation 
leading to ductile failure has been observed at high temperatures in the range 450-1000 K. At the intermediate temperature 
of 400 K, the nanowire yields through nucleation of crack associated with many mobile 1/2$<$111$>$ and immobile $<$100$>$ 
dislocations at the crack tip and fails in ductile manner. The ductile-brittle transition observed in $<$111$>$ BCC Fe 
nanowires is appropriately reflected in the stress-strain behavior and plastic strain at failure. The ductile-brittle 
transition increases with increasing nanowire size. The change in fracture behavior has been discussed in terms of the 
relative variations in yield and fracture stresses and change in slip behavior with respect to temperature. Further, the 
dislocation multiplication mechanism assisted by the kink nucleation from the nanowire surface observed at high temperatures 
has been presented. \\ 


\end{abstract}

%\end{onecolabstract}
\maketitle
%\doublespacing
%]
%\renewcommand{\thefootnote}{\fnsymbol{footnote}} \footnotetext{* email :
%sg@igcar.gov.in}

\section{Introduction}

In recent years, one dimensional metallic nanowires have attracted a considerable attention for research due to their unique 
properties and potential applications in future nano/micro electro-mechanical systems (NEMS/MEMS)\cite{Lieber}. Due to the high 
surface area to volume ratio, the nanowires exhibit superior electrical, optical, and mechanical properties compared to their 
bulk counterparts. Particularly, BCC Fe nanowires, which possess good magnetic properties, find applications in data/memory 
storage devices, permanent magnets, spin electronics, magnetic field sensing, enhancement agents for magnetic resonance imaging 
(MRI), medical sensors, and other smart devices \cite{Appl-1,Appl-2,Appl-3}. Improving the durability and reliability of these 
devices requires fundamental understanding of mechanical properties and associated deformation mechanisms. With detailed 
understanding of deformation and failure behavior, the mechanical properties of nanowires can be significantly improved by 
a suitable micro-structural design \cite{angstrom-scale-twins}. Further, the knowledge of deformation behavior also becomes 
important for fine-tuning the physical properties of nanowires with respect to reorientation, shape-memory, pseudo-elasticity, 
and super-plasticity.

Designing and performing mechanical testing of nanowires at different sample sizes, temperatures, and strain rates generally 
require sophisticated testing methods. In this context, molecular dynamics (MD) simulations are known to play a key role in 
exploring the deformation behavior of metallic nanowires. Based on detailed MD simulations, it has been shown that BCC Fe 
nanowires deform mainly either by the full dislocation slip or by the twinning mechanism. The deformation behavior of FCC and 
BCC nanowires has been reviewed in detail in Refs. \cite{Cai-Review} and \cite{Weinberger-Review}. Under tensile loading, BCC 
Fe nanowires oriented in $<$100$>$, $<$112$>$ and $<$102$>$ axial directions deform predominantly by the twinning mechanism, 
whereas the full dislocation slip has been observed in $<$110$>$ and $<$111$>$ orientations \cite{Sai-CMS16}. Further, it has
been shown that BCC Fe nanowires exhibit tension-compression asymmetry in deformation mechanisms \cite{Sai-CMS16,Healy15}. In
contrast to tensile loading, the nanowire with $<$100$>$ orientation deformed by the dislocation slip, whereas twinning was
observed in $<$110$>$ orientation under compressive loading \cite{Sai-CMS16,Healy15,Thaulow}. Recently, Wang et al. \cite{Wang-Nature} 
have provided the first experimental evidence of deformation twinning in BCC W nanowires with 15 nm diameter thereby conforming 
the predictions made using MD simulations. In addition to perfect nanowires, deformation by twinning is also observed in surface 
coated Fe \cite{Oxidation} and Fe-Cr \cite{Alloy-NWs} nanowires. It is interesting to note that when the BCC Fe nanowire deforms 
by the twinning mechanism, it has been found to undergo reorientation \cite{Sai-CMS15,Reorientation} and exhibit the shape memory 
effect and pseudo-elasticity \cite{Reorientation,Cao,Li-PRB,Interface}. These mechanisms cannot be observed when the nanowires 
deform by the dislocation slip. The reorientation mechanism observed in BCC Fe nanowires exhibits a strong size and temperature 
dependence \cite{Sai-CMS15,Cao}. It has been shown that the deformation by twinning leads to reorientation in nanowires of cross 
section width less than 11.42 nm. Beyond this size even though nanowires still deform by the twinning mechanism, the reorientation 
mechanism ceases to operate due to twin-twin interactions \cite{Sai-CMS15}. In addition to twinning, BCC Fe nanowires also exhibit 
peculiar dislocation behavior with respect to nanowire size, temperature, and strain rate. It has been shown that during early 
plastic deformation, dislocation loops initially consist of edge as well as screw components. However, with increasing strain, 
the edge components escape to the surface due to higher mobility and this leads to the accumulation of straight screw dislocations 
in $<$110$>$ BCC Fe nanowires \cite{Sai-MSEA15}. During compressive deformation of $<$100$>$ BCC Fe nanowires, it has been shown 
that dislocation-mediated plasticity dominates at 500 K, whereas the twinning mechanism has been observed in addition to the 
dislocation slip at lower temperatures \cite{ADutta}. A similar behavior has been observed with respect to temperature in $<$110$>$ 
BCC Fe nanowires \cite{Thaulow}.

Most of these studies in BCC metallic nanowires have been focused mainly on deformation aspects such as twinning and dislocation 
behavior \cite{Sai-CMS16,Healy15,Thaulow,Wang-Nature,Oxidation,Alloy-NWs,Sai-CMS15,Reorientation,Cao,Li-PRB,Interface,Sai-MSEA15,
ADutta}. It is surprising that enough attention has not been paid towards the fracture behavior of BCC Fe nanowires. It is well 
known that BCC materials are generally difficult to deform at low temperatures leading to brittle fracture, whereas at high temperature, 
they fail by ductile manner. At nanoscale, the brittle to ductile transition with respect to temperature has been observed in 
semiconductor nanowires such as GaN \cite{GaN}, ZnO \cite{ZnO}, and Si \cite{Pizzagalli-MSMSE}. Similar transition has also 
been reported in Cu nanowires with respect to size \cite{Cu-DBTT} and Ag nanowires with respect to the strain rate \cite{Ag-DBTT}. 
It has been reported that the pre-cracked bulk single crystal BCC Fe oriented in the $<$100$>$/\{100\} direction exhibits brittle 
to ductile transition with increasing temperature \cite{BCC-Fe-DBT}. The temperature for brittle to ductile transition in the bulk
BCC Fe single crystal has been found to vary in the range 130-154 K depending on the imposed strain rate. In addition to single crystal 
BCC Fe, Fe-9\%Cr and Fe based ferritic steels also exhibit the brittle to ductile transition \cite{BCC-Fe-DBT-2}. In this context, 
it is important to examine whether there is a similar brittle to ductile transition in BCC Fe at nanoscale? In view of this, detailed 
MD simulations were performed on the tensile deformation of $<$111$>$ BCC Fe nanowires at temperatures ranging from 10 to 1000 K in 
the present study. Emphasis has been given on the nature of defect nucleation with respect to temperature, the amount of plastic 
deformation before failure, and fracture behavior.

\section{MD Simulation Details}

MD simulations have been performed in Large scale Atomic/Molecular Massively Parallel Simulator (LAMMPS) package \cite{LAMMPS} 
employing an embedded atom method (EAM) potential for BCC Fe given by Mendelev and co-workers \cite{Mendelev}. The important 
outcome of MD simulations depends solely on the reliability of inter-atomic potential employed. It has been observed that some 
EAM potentials are reliable for many FCC systems, whereas for BCC systems, they may be less accurate in reproducing the defect 
structures, slip systems, twinning behavior, and phase transition. This difference in behavior with respect to the crystal 
structure results mainly from the non-planar core of screw dislocations in BCC systems, which makes the slip more complex compared 
to FCC systems. In the present investigation, the Mendelev EAM potential has been chosen mainly because several predictions made 
with this potential are in good agreement with either experimental observations or density functional theory (DFT) calculations. 
In agreement with DFT calculations \cite{core-DFT}, Mendelev EAM potential predicts a non-degenerate core structure for screw 
dislocations \cite{non-degenerate}. In contrast, all other potentials for BCC Fe predict a degenerate core structure. Similarly, 
depending on orientation, this potential correctly predicts deformation by twinning and dislocation slip in BCC Fe nanowires 
\cite{Sai-CMS16,Healy15,Sai-CMS15,Sai-MSEA15,ADutta}, which is quite close to the recent experimental observations in ultra-thin 
BCC W nanopillars \cite{Wang-Nature}. The mechanism of twin nucleation and growth \cite{Sai-CMS15,Sai-PhilMag16}, twin boundary 
as a dislocation source \cite{Sai-PhilMag16,ADutta}, twin migration stress \cite{Sai-CMS15,Ohja-PhilMag}, twist boundary structure 
\cite{Sai-PhilMagLett}, accumulation of straight screw dislocations \cite{Sai-MSEA15}, and various twin-twin interactions 
\cite{Sai-PhilMag16,Ohja-PhilMag} are in good agreement with those observed experimentally \cite{Ohja-PhilMag,Paxton,Hull,Ohr-TWGB}.

BCC Fe nanowires of square cross-section width (d) = 8.5 nm and oriented in $<$111$>$ axial direction with \{110\} and \{112\} 
side surfaces were created by generating atomic positions corresponding to the bulk Fe. Correspondingly, the simulation box contained 
about 110000 Fe atoms arranged in BCC lattice. The length (l) was twice the cross section width (d) of the nanowire. Periodic boundary 
conditions were chosen along the length direction, whereas the other two directions were kept free in order to mimic an infinitely 
long nanowire. After the initial construction of the nanowire, the energy minimization was performed by a conjugate gradient (CG) 
method to obtain a relaxed structure with equilibrium atomic positions corresponding to the nanowire. The minimization has been 
carried out until the energy change between two successive iterations divided by the initial energy is less than $10^{-6}$. To put 
the sample at the required temperature, all the atoms have been assigned initial velocities according to the Gaussian distribution. 
Following this, the nanowire system was thermally equilibrated to a required temperature for 125 ps in canonical ensemble (constant 
NVT). The temperature is controlled with Nose-Hoover thermostat with a damping constant of 500 fs for all temperatures. This value 
of damping constant has ensured that the temperature fluctuations are always lower than 1\% during simulation irrespective of test 
temperature. The velocity verlet algorithm has been used to integrate the equations of motion with a time step of 5 fs. In order to 
examine the influence of time step, the simulations have also been carried out with a lower time step of 3 fs. It has been observed 
that there are no significant differences in the observed results.

Following thermal equilibration, the tensile deformation was performed at a constant strain rate of $1\times$ $10^8$ $s^{-1}$ along 
the axis of the nanowire. The strain rate considered here is higher than the typical experimental strain rates due to the inherent 
time-scale limitations in MD simulations. In order to reveal the brittle to ductile transition temperature, the atomistic simulations 
have been performed at different temperatures in the range 10 - 1000 K. The simulations at each temperature have been repeated five 
times, with a different random number seed for velocity distribution each time. The stress was calculated from the Virial definition 
of stress \cite{Virial-Yip,Virial-Zhou} which takes the form

$$ \sigma_{\alpha \beta} = \frac{1}{V} \Bigg[ \frac{1}{2} \sum_{i}^N \sum_{j\neq 1}^{N} F_{ij}^{\alpha} r_{ij}^{\beta} - 
\sum_{i}^N m_i v_i^\alpha v_i^\beta \Bigg] $$

where N is the total number of atoms, V is the volume of the nanowire or simulation cell, $r_{ij}$ is the distance between atoms i 
and j, $F_{ij}$ is the force between atoms i and j, $m_i$ and $v_i$ are the mass and velocity of particle i, and the indices $\alpha$ 
and $\beta$ denote the Cartesian components. The first term in the above equation is due to the inter-atomic force and the second 
term is due to thermal vibrations, which becomes important with increasing temperature. The strain has been obtained as $(l-l_0)/l_0$, 
where $l$ is the instantaneous length of the nanowire and $l_0$ is the initial length. The visualization of atomic configurations was 
performed using OVITO \cite{Ovito}. 

\section{Results}
\subsection{Stress-strain behavior}

\begin{figure}
\centering
\includegraphics[width= 8.5cm]{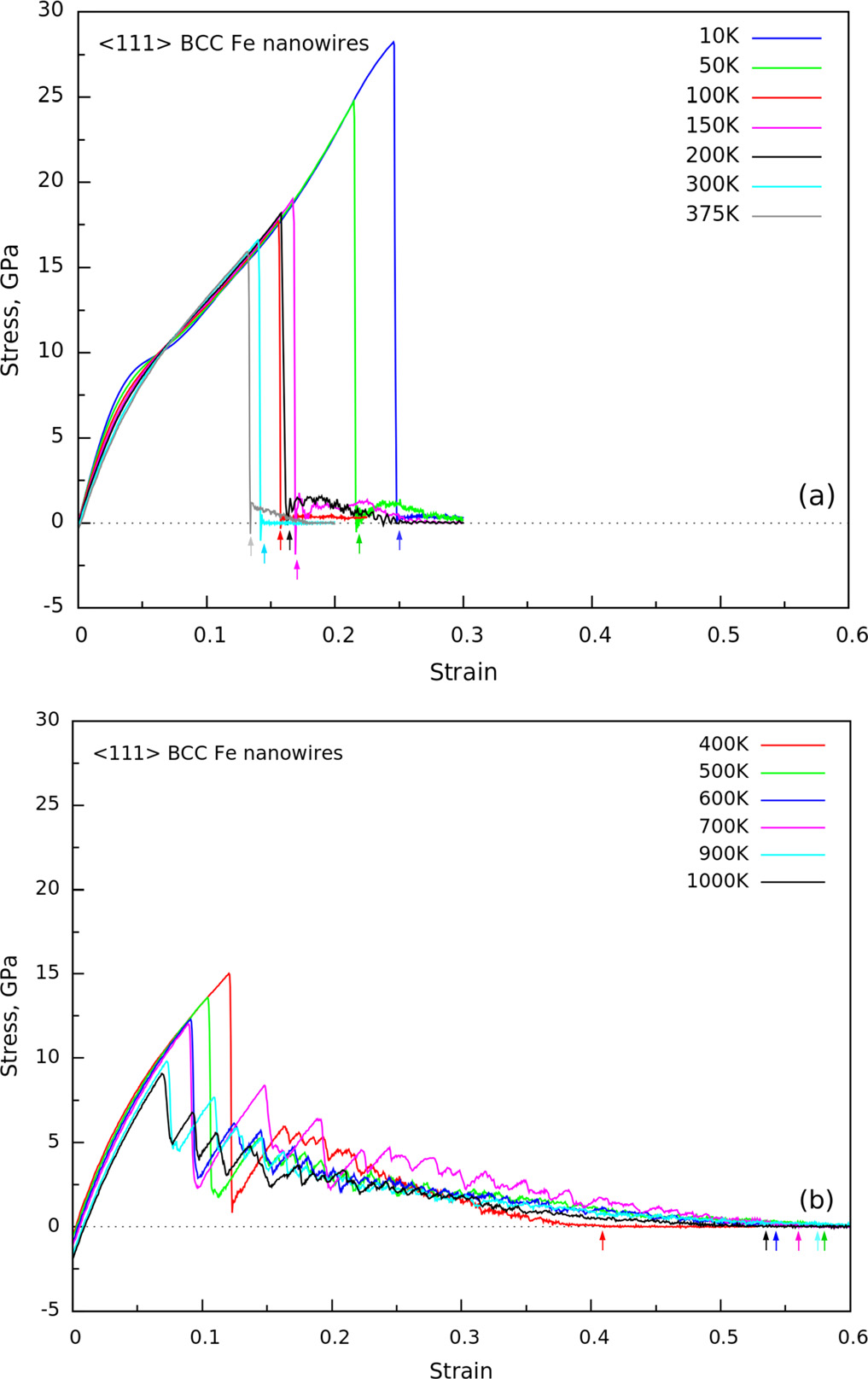}
\caption { Stress-strain behavior of $<$111$>$ BCC Fe nanowires at (a) low (10-375 K) and (b) high (400-1000 K) 
temperatures. The failure locations have been marked by arrows. Since the loading in the present investigation is in the z -direction, 
the stress here refers to the z-component of the stress tensor, i.e., $\sigma_{zz}$}
\label{stress-strain}
\end{figure}

Figure \ref{stress-strain} shows the stress-strain behavior of $<$111$>$ BCC Fe nanowires at various temperatures ranging from 10 to
1000 K. It can be seen that all the nanowires undergo an elastic deformation and exhibit non-linear behavior at high strains. Further, 
during the elastic deformation at 10 and 50 K, a bump appears in the stress-strain curve at a strain value of 0.4-0.5. This bump may 
be due to weird behavior of surface atoms at low temperatures or it may arise from the inter-atomic potential. Following elastic 
deformation up to peak stress, yielding results in abrupt drop in flow stress in all the nanowires. The yield stress values have been 
obtained as the values of peak stress in the stress-strain curve. Following yielding, the $<$111$>$ BCC Fe nanowires display two 
different behaviors during plastic deformation as shown for low (10-375 K) and high (400-1000 K) temperatures in Figs. \ref{stress-strain}(a) 
and \ref{stress-strain}(b), respectively. At low temperatures, the flow stress abruptly drops to zero indicating insignificant plastic 
deformation and sudden failure in the nanowires (Fig. \ref{stress-strain}(a)). On the other hand, the flow stress dropping to nonzero 
value followed by jerky flow and gradual decrease in flow stress during plastic deformation suggests ductile nature of nanowires 
at high temperatures (Fig. \ref{stress-strain}(b)). 

\begin{figure}
\centering
\includegraphics[width= 8.5cm]{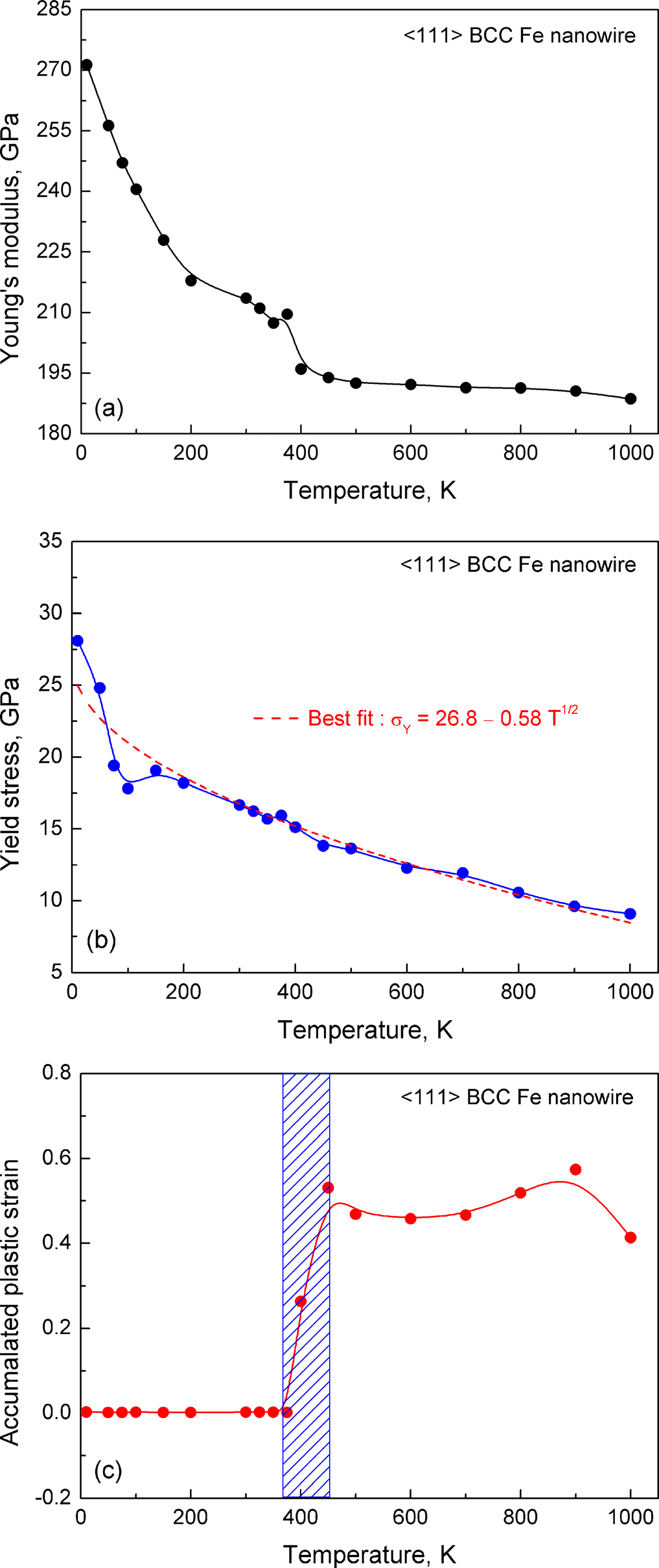}
\caption { Variations of (a) Young’s modulus, (b) yield stress, and (c) accumulated plastic strain as a function of 
temperature for $<$111$>$ BCC Fe nanowires. Temperature (T) dependence of yield stress ($\sigma_Y$) obeying $\sigma_Y = A - B\sqrt{T}$ 
with A = 26.8 and B = 0.58 is superimposed as the broken line in (b). The blue dashed region in (c) shows a ductile to brittle transition 
regime. The center of this regime has been taken as the transition temperature.}
\label{Properties}
\end{figure}

The variations in Young’s modulus, yield stress, and the accumulated plastic strain before failure as function of temperature are 
shown in Figs. \ref{Properties}(a)-\ref{Properties}(c). In the presence of non-linear elastic deformation, the values of Young’s 
modulus at different temperatures have been evaluated from the slope of initial linear elastic regime, that is, the slope of 
stress-strain curves for $\varepsilon < 0.04$. Below this strain, the stress-strain curve is nearly linear for all temperatures. 
The non-linear portion at high elastic strains ($\varepsilon > 0.04$) has been neglected for Young’s modulus calculations. It can 
be seen that Young’s modulus decreases rapidly up to 400 K followed by saturation at higher temperatures (Fig. \ref{Properties}(a)). 
The variations in yield stress with respect to temperature exhibited a rapid decrease in yield stress up to 100 K followed by a 
gradual decrease with the increase in temperature (Fig. \ref{Properties}(b)). At about 100 K, a hump in yield stress or a concave 
down region can be seen in Fig. \ref{Properties}(b). Temperature (T) dependence of yield stress ($\sigma_Y$) obeying $\sigma_Y = 
A - B\sqrt{T}$ with A = 26.8 and B = 0.58 is superimposed as the broken line in Fig. \ref{Properties}(b). The yield stress values 
at all temperatures have been observed to be consistently lower than the theoretical strength of 27.6 GPa reported for $<$111$>$ 
BCC Fe \cite{Strength-Friak}. It is interesting to observe that apart from insignificant flow stress (Fig. \ref{stress-strain}(a)), 
the nanowires exhibit a negligible plastic strain during tensile deformation at low temperatures in the range 10-375 K (Fig. 
\ref{Properties}(c)). A significant increase in the accumulated plastic strain with the increase in temperature from 375 to 500 K 
can be seen in Fig. \ref{Properties}(c). Beyond 500 K, the plastic strain remains nearly constant. The accumulated plastic strain 
is obtained from stress-strain curves as total strain to failure minus elastic strain (the yield strain). The negligible amount of 
accumulated plastic strain obtained at lower temperatures arises from crack nucleation and growth till separation into two pieces. 
At higher temperatures, significant plastic deformation leading to failure results in higher values of accumulated plastic strain. 
From the variations in accumulated plastic strain before failure with temperature, it is clear that $<$111$>$ BCC Fe nanowires 
undergo ductile-brittle transition with 400 K as a transition temperature.

\subsection{Deformation and failure behavior}

\begin{figure}
\centering
\includegraphics[width= 8.5cm]{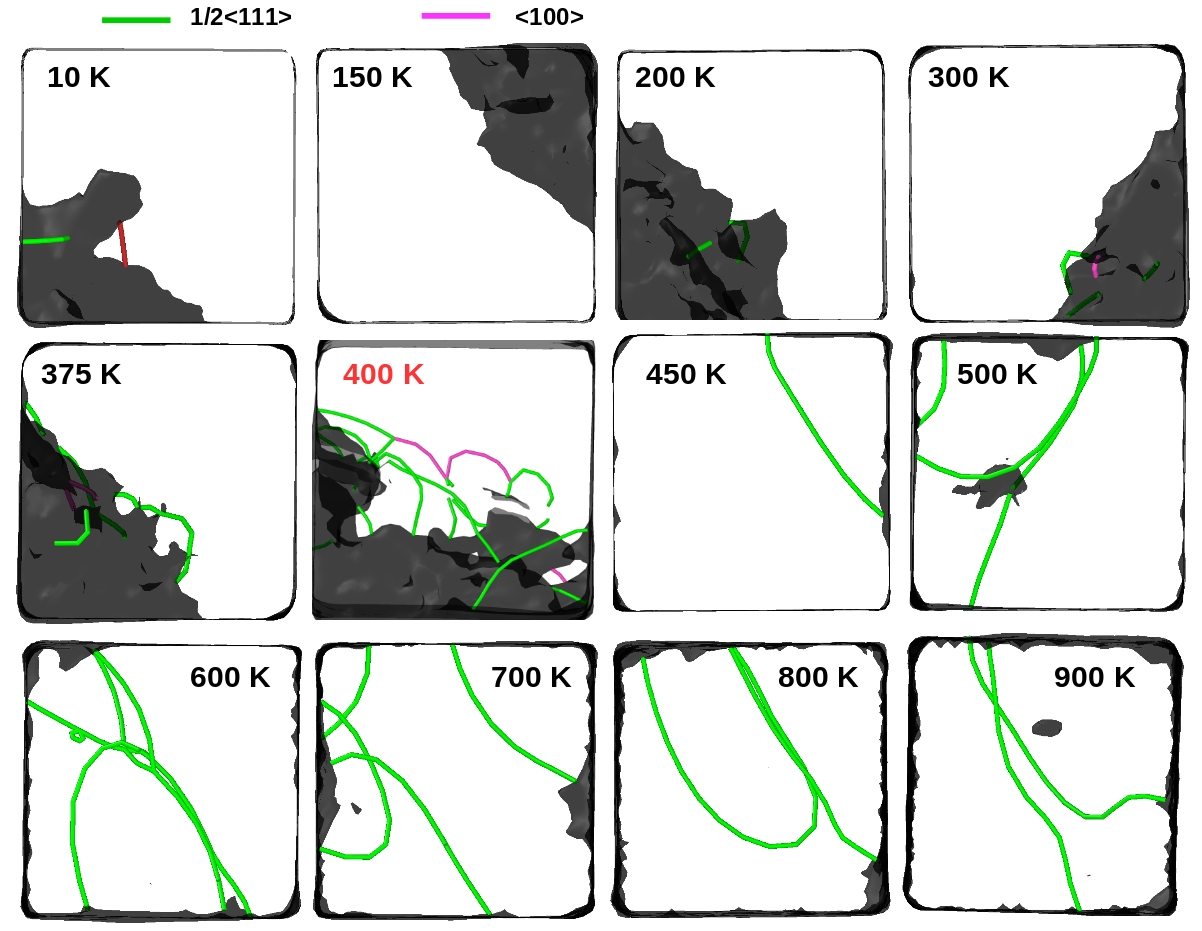}
\caption { Defect nucleation at yielding in $<$111$>$ BCC Fe nanowires at different temperatures. The green lines indicate
the dislocations with the Burgers vector 1/2$<$111$>$, the magenta lines represent the dislocations with the Burgers vector $<$100$>$, 
and the dislocations with the unknown/unidentified Burgers vector are shown by red lines. The black regions indicate defective surfaces
such as cracks.}
\label{Nucleation}
\end{figure}

\begin{figure}
\centering
\includegraphics[width= 8.5cm]{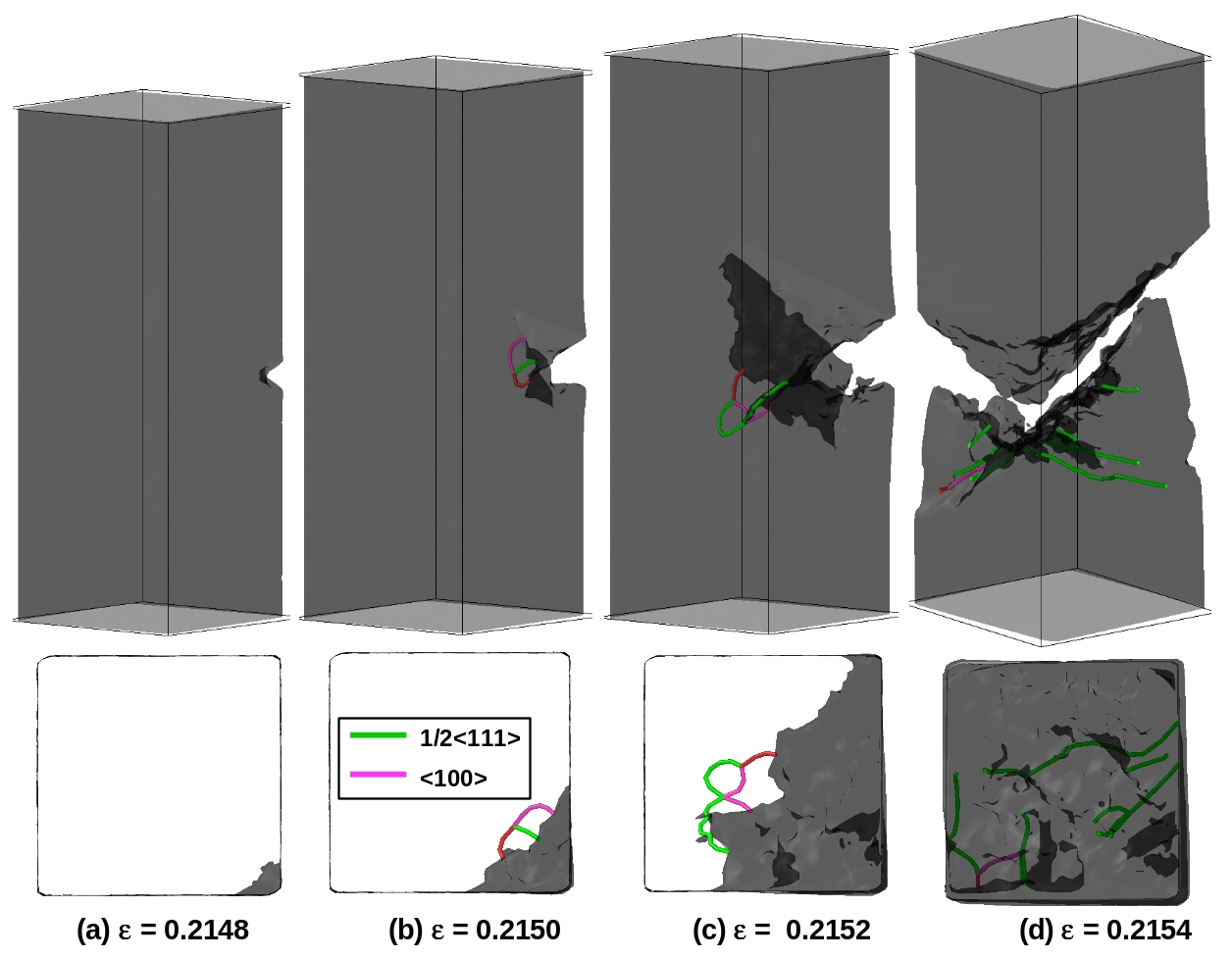}
\caption { Atomic snapshots as a function of total strain at 50 K representing deformation behavior of $<$111$>$ Fe 
nanowires at low temperatures in the range 10-375 K. The color code details are described in Fig. \ref{Nucleation} caption.}
\label{Low-T-deform}
\end{figure}

In order to understand the variations in stress-strain behavior, the atomic configurations have been analyzed using OVITO as a function 
of strain at different temperatures. Figure \ref{Nucleation} shows the top view of nanowires during yielding at various temperatures. 
It can be seen that at low temperatures (10-375 K), yielding in the nanowires occurs mainly through the nucleation of a sharp crack 
with negligible dislocation activity at the crack tip. At transition temperature of 400 K, the nanowire yields primarily through the
nucleation of crack associated with many mobile 1/2$<$111$>$ and immobile $<$100$>$ dislocations in the vicinity of the crack. At 
temperatures higher than 400 K, the nanowires yielded only by the nucleation of multiple 1/2$<$111$>$ dislocations. In the temperature 
range 450-1000 K, no crack nucleation has been observed. The atomic snapshots as function of strain at 50 K representing deformation 
behavior of $<$111$>$ BCC Fe nanowires at low temperatures is shown in Fig. \ref{Low-T-deform}. It can be seen that the crack nucleates 
from the corner of the nanowires without any dislocations at the tip (Fig. \ref{Low-T-deform}(a)) and grows rapidly along the direction 
at $45^o$ angle with the loading axis (Figs. \ref{Low-T-deform}(b) and \ref{Low-T-deform}(c)). During crack growth, a few 1/2$<$111$>$ 
and $<$100$>$ dislocations in the vicinity of crack tip can be seen in Figs. \ref{Low-T-deform}(b) and \ref{Low-T-deform}(c). The growing 
crack reaches the opposite surface within a short strain interval and the nanowire fails abruptly without showing any plastic deformation 
(Fig. \ref{Low-T-deform}). These observations clearly suggest that $<$111$>$ BCC Fe nanowires fails in brittle manner in the temperature
range 10-350 K and the peak stress in the stress-strain curves reflects the fracture strength of the nanowires. At 400 K, apart from 
yielding through nucleation of crack, several mobile 1/2$<$111$>$ and immobile $<$100$>$ dislocations near crack tip have been observed 
(Fig. \ref{Nucleation}). With increasing deformation, the crack gets blunted by dislocation activity and as a result, the nanowire at 
400 K exhibits considerable plastic deformation and fails in ductile manner at high strains.

\begin{figure}
\centering
\includegraphics[width= 8.5cm]{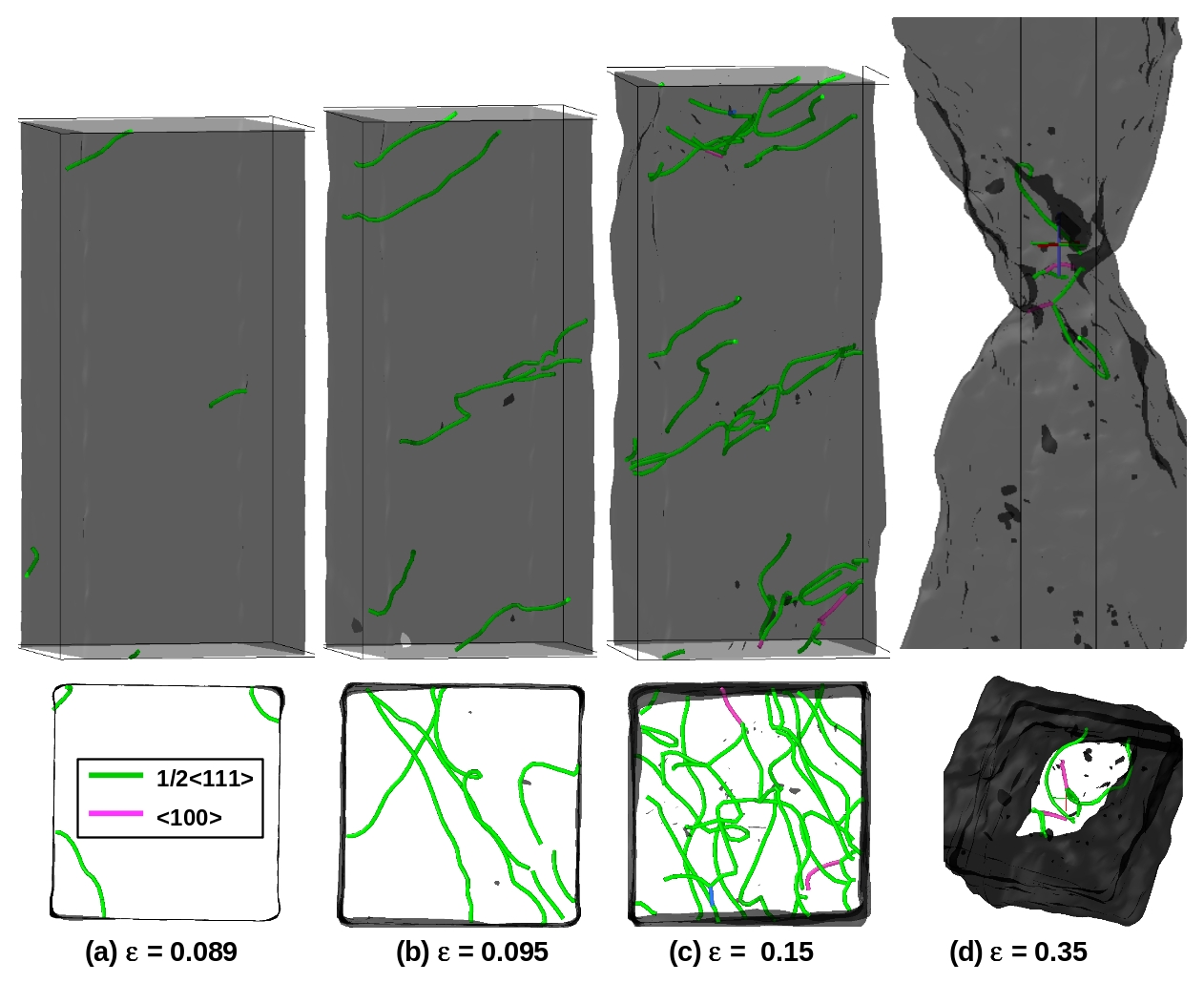}
\caption { Atomic snapshots as a function of total strain at 700 K representing deformation behavior of $<$111$>$ Fe
nanowires at high temperatures in the range 450-1000 K. The color code details are described in Fig. \ref{Nucleation} caption.}
\label{High-T-deform}
\end{figure}

\begin{figure}
\centering
\includegraphics[width= 8.5cm]{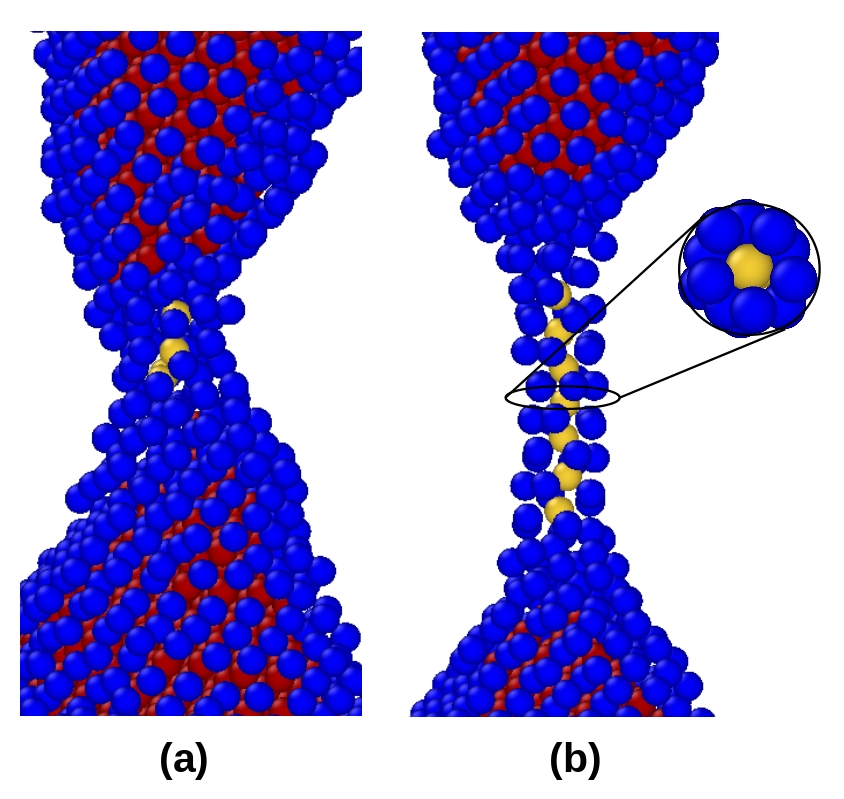}
\caption { Formation of pentagonal atomic chain in the necking region during deformation of $<$111$>$ Fe nanowires at 
900 K. The red color represents perfect BCC atoms, blue color indicates the atoms in the non-crystalline structure, and yellow 
indicates the atoms in five fold symmetry.}
\label{Pentagonal}
\end{figure}

Contrary to crack nucleation at low temperatures, the nanowires yield through the nucleation of multiple 1/2$<$111$>$ dislocations 
at high temperatures in the range 450-1000 K (Fig. \ref{Nucleation}). Therefore, the peak stresses in the stress-strain curves at 
high temperatures necessarily indicate stress for the nucleation of dislocations in an otherwise perfect nanowire. At high 
temperatures, yielding through dislocation nucleation followed by plastic deformation results in flow stress drop to non-zero 
values (Fig. \ref{stress-strain}(b)). The atomic configurations representing typical yielding and plastic deformation of nanowires 
at high temperatures (450-1000 K) are shown for 700 K in Fig. \ref{High-T-deform}. The yielding by the nucleation of dislocations 
from the nanowire corner can be seen in Fig. \ref{High-T-deform}(a). It can also be seen that the dislocations nucleated from the 
corners glide with the increase in plastic deformation on their respective glide planes (mainly \{110\} type) and eventually escape 
to the surface (Figs. \ref{High-T-deform}(b) and \ref{High-T-deform}(c)). The continuous nucleation and glide of dislocations on 
interacting glide planes leads to the formation of well defined necking (Fig. \ref{High-T-deform}(d)) and the nanowires fail in 
ductile manner at significantly higher plastic strains. Interestingly at high temperatures of 900 and 1000 K, it has been observed 
that the disordered atoms in the neck region rearrange themselves and forms the pentagonal atomic chain as shown in Fig. \ref{Pentagonal}. 
The atomic chain consists of a central atom sandwiched between the two pentagonal rings, where each pentagonal ring consists of 5
atoms. Formation of pentagonal atomic chains has been reported in $<$100$>$ BCC Fe and Fe-Cr nanowires at high temperatures for ultra 
thin cross-section width less than 2.83 nm \cite{Alloy-NWs,Sai-MRX}. The transformation of perfect BCC lattice into the pentagonal 
structure has been described in terms of energy minimization \cite{Sai-MRX,Pentagonal-PRL}.

\subsection{Dislocation multiplication mechanism and dissociation of immobile dislocation}

\begin{figure}
\centering
\includegraphics[width= 8.5cm]{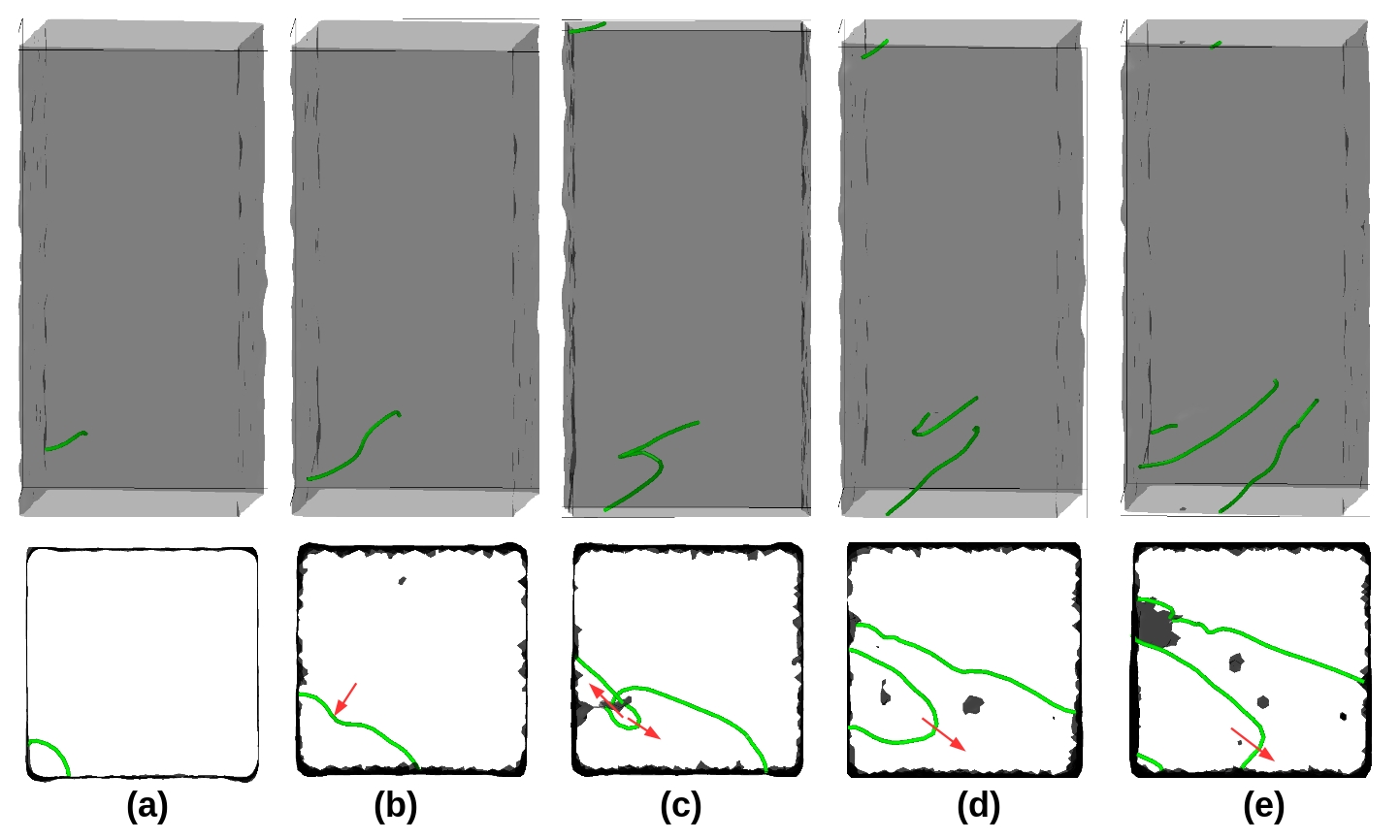}
\caption {A special dislocation multiplication mechanism observed in $<$111$>$ Fe nanowires at high temperatures in the range 450–1000 K. 
The color code details are described in Fig. \ref{Nucleation} caption.}
\label{Multiplication}
\end{figure}

In bulk materials, the dislocations generally multiply by the well known Frank-Read mechanism. However, the Frank-read mechanism 
no longer operates when the size is reduced to nanoscale. A special dislocation multiplication mechanism operating during plastic 
deformation of $<$111$>$ BCC Fe nanowires at high temperatures (450–1000 K) is shown in Fig. \ref{Multiplication}. It can be seen 
that when a mixed dislocation nucleates from the nanowire corner (Fig. \ref{Multiplication}(a)), it align itself to a straight 
screw configuration with further glide (Fig. \ref{Multiplication}(b)). This straight screw dislocation glides through the kink-pair 
mechanism, where the kinks nucleate from the nanowire surfaces. When the kinks having different orientations nucleated from two 
different surfaces move towards each other, a cusp develops at their intersection point on the straight screw dislocation (Figs. 
\ref{Multiplication}(b) and \ref{Multiplication}(c)). With increasing deformation, the radius of curvature of the cusp decreases 
(as observed from the nanowire axial direction, i.e., top view) and it appears like a dislocation loop as shown in Fig. 
\ref{Multiplication}(c). As the loop grows with strain, the edge component of the loop reaches the nearby nanowire surface and this 
creates three independent screw dislocations in the nanowire (Figs. \ref{Multiplication}(d) and \ref{Multiplication}(e)). The 
mechanism of kink nucleation and glide from the nanowire surface resulting in dislocation multiplication into three independent 
dislocations has been shown in Mo nanopillars by Weinberger and Cai \cite{self-multiplication} using a combination of MD and 
dislocation dynamics simulations. It has been suggested that a similar dislocation multiplication will be operating in other 
BCC nanowires \cite{Cai-Review} as observed in the present study. This observation further supports the applicability of Mendelev 
EAM potential for understanding deformation behavior in BCC Fe nanowires. For the operation of this special dislocation multiplication 
mechanism, Lee et al. \cite{Lee-Woo} laid down necessary conditions on the mobility of edge and screw dislocations. Using dislocation 
dynamics simulations, it has been shown that the surface controlled dislocation multiplication occurs only when the screw dislocation
mobility is much lower than that of edge dislocation \cite{Lee-Woo}. When the mobility of edge and screw dislocations is equal, the 
complete dislocation gets annihilated at the surface and the dislocation multiplication does not occur \cite{Lee-Woo}.

\begin{figure}
\centering
\includegraphics[width= 8.5cm]{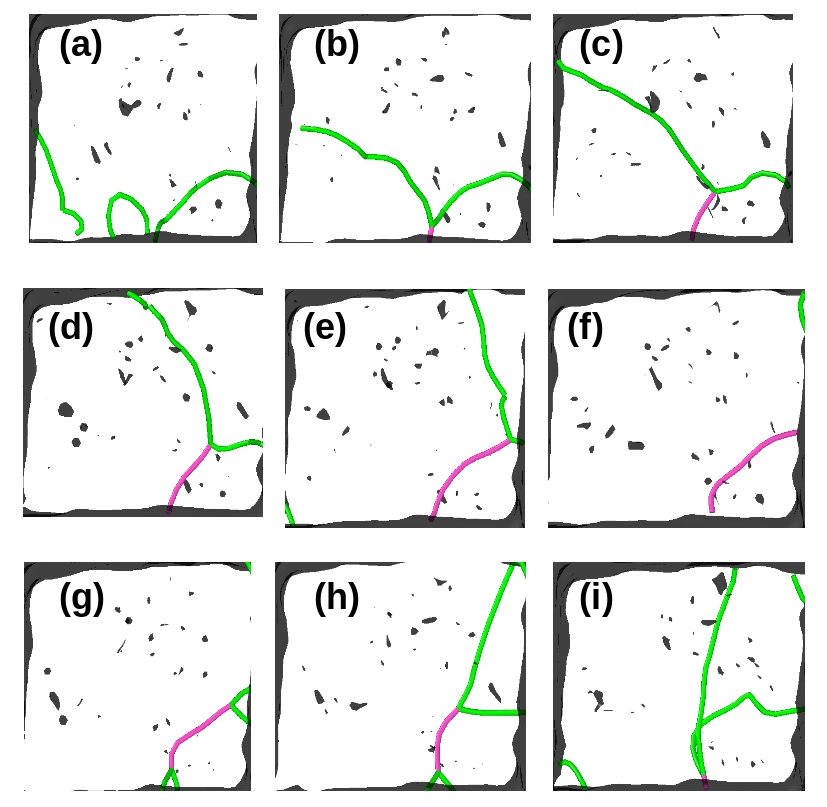}
\caption {Surface assisted formation and dissociation of $<$100$>$ immobile dislocations through two 1/2$<$111$>$ mobile 
dislocations at high temperatures. The color code details are described in Fig. \ref{Nucleation} caption.}
\label{Immobile}
\end{figure}

It has been observed that the deformation behavior of $<$111$>$ BCC Fe nanowires is dominated by the slip of 1/2$<$111$>$ mobile 
dislocations at high temperatures (450-1000 K). In addition to 1/2$<$111$>$ dislocations, $<$100$>$ immobile dislocations have 
also been observed. The different stages of the formation and dissociation of $<$100$>$ immobile dislocation emanating from the 
interactions of two 1/2$<$111$>$ mobile dislocations are shown in Fig. \ref{Immobile}. Initially, two 1/2$<$111$>$ dislocations 
nucleate from the nanowire surface and attract towards each other (Fig. \ref{Immobile}(a)). With increasing strain, part
of these two dislocations combine and form a $<$100$>$ immobile dislocation (Fig. \ref{Immobile}(b)). Subsequent glide of these
mobile dislocations at Y-junction leads to a zipping process, which increases the length of the immobile dislocation (Figs. 
\ref{Immobile}(c) and \ref{Immobile}(d)). Following completion of zipping process, long $<$100$>$ immobile dislocation is obtained
(Figs. \ref{Immobile}(e) and \ref{Immobile}(f)). The formation of $<$100$>$ dislocation has also been observed during the compressive 
deformation of $<$100$>$ BCC Fe nanopillars \cite{ADutta}. Generally, it is assumed that $<$100$>$ dislocation is highly stable and 
immobile and aids in the nucleation of micro-cracks in BCC metals. Contrary to this, it has been observed in the present study
that $<$100$>$ dislocation is not stable during the deformation of nanowires and dissociates into two mobile dislocations with 
Burgers vector 1/2$<$111$>$ as shown in Figs. \ref{Immobile}(g)-\ref{Immobile}(i). The dissociation/unzipping process of $<$100$>$ 
dislocations initiates from the surface of the nanowire through the formation of two Y-junctions (Fig. \ref{Immobile}(g)) and with 
increasing deformation, one of the Y-junction penetrates towards the other end of the dislocation (Figs. \ref{Immobile}(h) and 
\ref{Immobile}(i)). According to Frank’s criterion, the dissociation of $<$100$>$ dislocation is difficult to be observed as this 
reaction leads to the increase in energy (i.e., $b_1^2 < b_2^2 + b_3^2$, where $b_1$ is the Burgers vector of immobile dislocation 
and $b_2$ and $b_3$ are the Burgers vector of dissociated dislocations). However, this dislocation reaction becomes feasible at 
high energy or stresses in the order of GPa as observed in the present study. Since the nanowire surfaces can act as local stress
raisers, it is reasonable to conclude that the surfaces of the nanowires/nanopillars provide a source for the dissociation of 
immobile dislocations in BCC nanowires. The dissociation of $<$100$>$ screw dislocation has also been observed at the twist 
boundary of BCC Fe and it has been shown that the surfaces aid to the dissociation \cite{Sai-PhilMagLett}.

\section{Discussion}

\subsection{Temperature dependence of yield stress}

In BCC/FCC metallic nanowires, the variations in yield stress ($\sigma_Y$) with temperature (T) follow either $\sigma_Y = A - BT$ 
or $\sigma_Y = A - B\sqrt{T}$ relation depending upon the inter-atomic potential used in MD simulations \cite{Cai-Review,Rabkin}. 
In BCC Fe nanowires, it has been shown that the yield stress follows $\sigma_Y = A - B\sqrt{T}$ relation with Mendelev EAM potential 
\cite{Kotrechko}. In agreement with above observation, the observed decrease in yield stress with the increase in temperature in the 
present study followed the relation  $\sigma_Y = 26.8 - 0.58\sqrt{T}$. It can be seen that the observed temperature dependence of 
yield stress in BCC Fe nanowires (Fig. \ref{Properties}(b)) is different from their bulk single crystal counterparts. In bulk single 
crystals, the yield stress generally saturates above a critical temperature \cite{MSEA-Hump,Diehl-Hump,Tangri-Hump,Kuramoto-Hump}. 
In BCC Fe nanowires, saturation in yield stress has not been observed. This difference in behavior essentially arises from the 
difference in yielding events in BCC Fe nanowires and bulk single crystals \cite{Kotrechko}. In nanowires, yielding results from the
nucleation of defects, whereas movement of existing dislocations leads to yielding in the bulk single crystals. The presence of a 
hump or concave down region in the yield stress - temperature curve at about 100 K (Fig. \ref{Properties}(b)) is in agreement with 
the experimental observations reported for the pure bulk single crystal of BCC Fe \cite{MSEA-Hump,Diehl-Hump,Tangri-Hump,Kuramoto-Hump}. 
However, the occurrence of hump in the bulk single crystal has been observed in the temperature range 200-250 K. Further, it has been 
shown that this hump disappears in Fe specimens doped with a small amount of carbon atoms \cite{MSEA-Hump,Diehl-Hump}. These studies 
indicate that the hump is intrinsic to pure BCC Fe lattice. Guyot and Dorn \cite{Guyot-Dorn} have suggested that this hump in flow 
stress is due to the double hump shape of Peierls potential in BCC lattice. Interestingly, the Mendelev EAM potential also shows the
double hump Peierls potential for BCC Fe \cite{Ventelon}. Therefore, the hump in yield stress versus temperature curve observed in
the present investigation can be attributed to the double hump shape of the Peierls potential.

\subsection{Ductile-brittle transition}

Influence of temperature on tensile deformation of$<$111$>$ BCC Fe nanowire clearly indicated that the nanowires yield through the 
nucleation of a sharp cracks and fail in brittle manner at low temperatures (10-375 K), whereas nucleation of multiple dislocations 
at yielding followed by significant plastic deformation leads to ductile failure at high temperatures (450-1000 K). At 400 K, the 
nanowire yielded by crack nucleation and reasonable dislocation activity at the crack tip resulted in ductile failure. The above 
failure behavior with respect to temperature is also reflected in the tensile ductility. BCC Fe nanowires displayed negligible plastic 
strain at low temperatures (10-375 K) followed by a rapid increase in the accumulated plastic strain in the temperature range 375-500 
K and high but nearly constant plastic strain above 500 K. These observations clearly suggest that BCC Fe nanowires display ductile-
brittle transition at 400 K. Similar ductile-brittle transition has been observed in bulk single crystal BCC Fe \cite{BCC-Fe-DBT}. 
However, the transition temperature of 400 K observed in nanowires is significantly higher than 130 K reported for the strain rate 
$4.46 \times 10^{-5}$ s$^{-1}$ in bulk single crystal BCC Fe \cite{BCC-Fe-DBT}. An increase in transition temperature to 154 K with 
the increase in strain rate to $4.46 \times 10^{-3}$ s$^{-1}$ has also been reported for bulk single crystal BCC Fe \cite{BCC-Fe-DBT}. 
Based on the detailed investigation, the ductile to brittle transition in BCC Fe has been successfully modeled as a function of strain 
rate \cite{Kashinath}. A transition temperature of 320 K has been predicted for the strain rate $1 \times 10^{3}$ s$^{-1}$ in BCC Fe 
\cite{Kashinath}. In order to examine the strain rate dependence of ductile-brittle transition in Fe nanowires, MD simulations have 
also been carried out at a strain rate of $1 \times 10^{9}$ s$^{-1}$. A transition temperature of 450 K observed for $1 \times 10^{9}$ 
s$^{-1}$ clearly indicates that the ductile-brittle transition temperature (DBTT) increases with increasing strain rate in the Fe 
nanowire. However, further studies are needed in this direction. The higher transition temperatures than the bulk counterparts observed 
in the present investigation can be ascribed as a consequence of high strain rates used in MD simulations.

\begin{figure}
\centering
\includegraphics[width= 8.5cm]{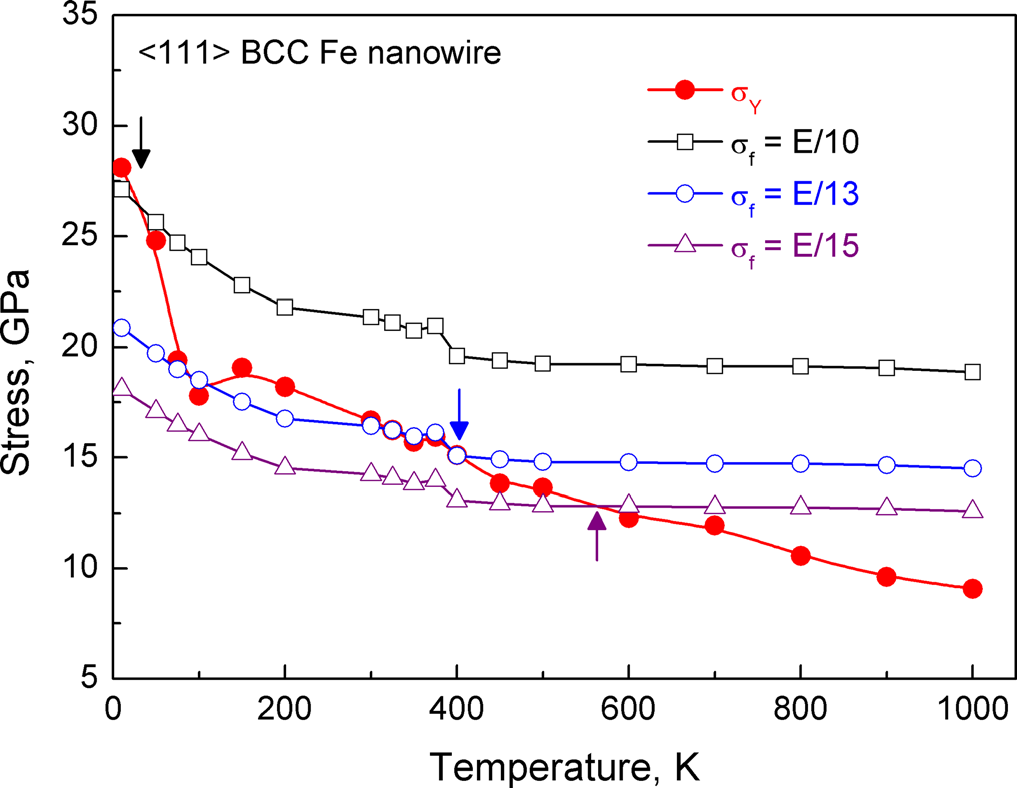}
\caption {Variations of yield stress as a function of temperature along with fracture stresses ($\sigma_f$) approximated to E/10, 
E/13, and E/15.}
\label{Predictions}
\end{figure}

The brittle to ductile transition behavior can be explained based on the relative variations of yield and fracture stresses with 
respect to temperature. It is known that Young’s modulus (E) is related to the perfect or fracture strength $\sigma_f$ of a material 
through the relation $\sigma_{f} = (E\gamma_s/a_0)^{1/2}$, where $\gamma_s$ is surface energy of the fractured surfaces and $a_0$ is 
interatomic spacing \cite{Dieter}. By making a reasonable approximation of $\gamma_s = Ea_0/20$, a rough estimate of strength in 
terms of Young’s modulus shows that $\sigma_f$ varies between E/4 and E/15 depending on material and test conditions \cite{Dieter}. 
The variations of yield strength and $\sigma_f$ (obtained by E/10, E/13, and E/15) as a function of temperature are shown in Fig. 
\ref{Predictions}. It can be clearly seen that $\sigma_f$ is less sensitive to temperature compared to yield stress and as a result, 
all the three $\sigma_f$ curves cross over yield strength at different temperatures. The temperatures of cross-over for E/10, E/13, 
and E/15 have been obtained as 40 K, 400 K, and 570 K, respectively. The temperature at which the fracture stress crosses the yield 
strength is considered as brittle-ductile transition temperature. Below the transition temperature, the fracture stress is either 
lower or close to yield stress and this leads to fracture before yielding. Above the transition temperature, fracture stress is much
higher than the yield stress and this result in yielding, significant plastic deformation, and ductile failure. From the above 
comparison, it is clear that $\sigma_f$ values evaluated as E/13 crosses the yield strength at 400 K as observed in the MD simulations. 
Therefore, $\sigma_f$ = E/13 can be assumed to represent the brittle to ductile transition behavior of $<$111$>$ BCC Fe nanowire.

\begin{figure}
\centering
\includegraphics[width= 8.5cm]{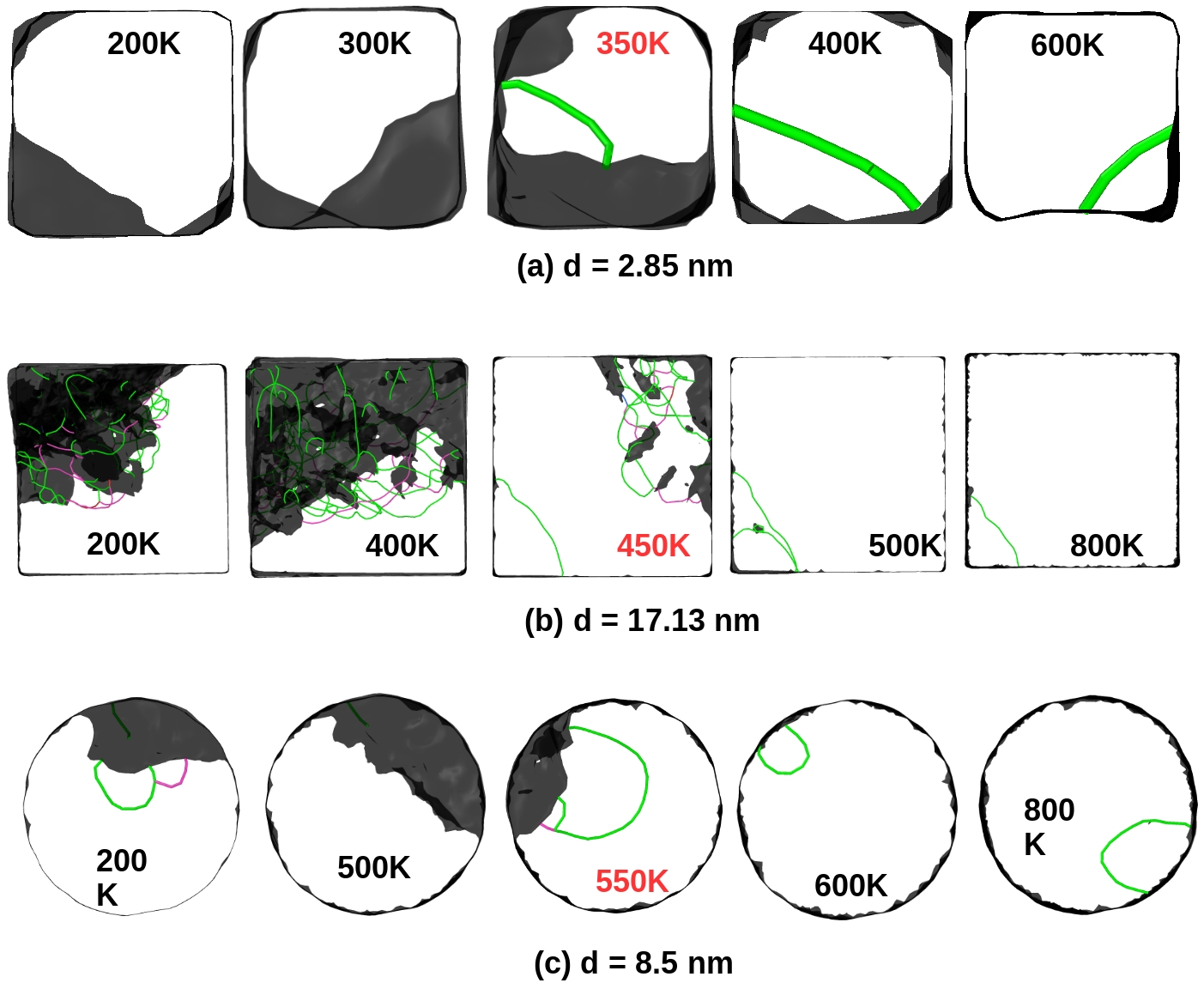}
\caption {Defect nucleation in square cross-section nanowires of (a) low and (b) high size with respect to temperature. Defect 
nucleation in the circular cross-section nanowire of d = 8.5 nm with respect to temperature is shown in (c). The ductile-brittle 
transition temperature in each case has been highlighted in the red color. The color code details are described in Fig. 
\ref{Nucleation} caption.}
\label{Size-shape}
\end{figure}

In a single phase BCC Fe-Co alloy, Johnston et al. \cite{Johnston} have shown that the change in fracture behavior is primarily
associated with a change in slip behavior and the yield stress plays a secondary role. In the BCC Fe-Co alloy, it has been shown 
that when the deformation is restricted to a planar glide, the alloy failed in brittle manner, whereas the wavy glide induces the 
ductile behavior \cite{Johnston}. This correlation has been ascribed to the requirement of at least five independent slip systems 
to induce a small and homogeneous strain, that is, von-Mises criterion. In the present study, it has been observed that when the 
nanowires failed in brittle manner, a few dislocations present at the crack tip have been associated with only one or two independent 
slip systems (Figure \ref{Low-T-deform}). On the other hand, when the nanowires show ductile behavior, the wavy glide is extensively 
observed as depicted in Figs. \ref{High-T-deform} and \ref{Multiplication}. In the case of wavy glide, the requirement of five 
independent slip systems is naturally met and this induces significant plastic deformation leading to ductile fracture. It is 
important to mention that the ductile-brittle transition observed in the $<$111$>$ BCC Fe nanowire in the present study is absent 
in $<$100$>$ and $<$110$>$ orientations of BCC Fe \cite{Sai-CMS15,Sai-MSEA15}. It has been observed that the BCC Fe nanowires
with $<$100$>$ and $<$110$>$ orientations undergo significant plastic deformation and fails via ductile mode even at the lowest
temperature of 10 K \cite{Sai-CMS15,Sai-MSEA15}. The activation of multiple slip systems satisfying the von-Mises criterion even 
at 10 K in $<$110$>$ BCC Fe nanowires lead to gross plastic deformation and high ductility \cite{Sai-MSEA15}. In $<$100$>$ BCC Fe 
nanowire, different deformation mechanisms of twinning and reorientation at 10 K results in high ductility \cite{Sai-CMS15}. Apart 
from orientation, the nanowire size and shape also play an important role on the deformation mechanisms. In order to reveal the 
influence of size on the observed ductile-brittle transition, further MD simulations have been carried out on two different nanowire 
sizes of 2.85 and 17.13 nm representing lower and higher sizes, respectively. The results indicate that the ductile-brittle transition 
temperature increases with increasing nanowire size. For the nanowire of cross-section width (d) = 2.85 nm, the transition has been 
observed at 350 K (Fig. \ref{Size-shape}(a)), whereas for nanowires with d = 17.13 nm, it has been observed at 450 K (Fig. \ref{Size-shape}
(b)). Similarly, the simulations performed with the circular cross-section nanowire of 8.5 nm diameter yields a transition temperature 
of 550 K (Fig. \ref{Size-shape}(c)), which is higher than that observed for the square cross-section nanowire. These results also 
indicate that the circular cross-section nanowires attains higher yield strength values, and as a result, the yield stress versus 
temperature curve in Fig. \ref{Predictions} crosses the fracture stress versus temperature curve at relatively higher temperature 
compared to the square cross-section nanowire.

\section{Conclusions}
The temperature dependent deformation and failure behavior of $<$111$>$ BCC Fe nanowires has been investigated by molecular dynamics 
simulations. The simulation results indicate that $<$111$>$ BCC Fe nanowires undergo ductile-brittle transition at 400 K. Below this 
temperature, the nanowires yield through the nucleation of a sharp crack and fails in a brittle manner, whereas at high temperatures, 
the nucleation of multiple 1/2$<$111$>$ dislocations associated with significant plastic deformation leads to ductile failure. The 
ductile-brittle transition in $<$111$>$ BCC Fe nanowires results from the relative variations of yield and fracture stresses as well 
as slip behavior with respect to temperature. Above the transition temperature of 400 K, the lower yield stress than the fracture 
stress facilitates yielding by the nucleation of dislocations and significant plastic deformation before ductile failure. This is well 
supported by the occurrence of wavy glide and dislocation multiplications at high temperatures. Below the transition temperature, the 
lower fracture stress than the yield stress leads to the yielding by crack nucleation and brittle failure with negligible dislocation 
activity in the nanowires. Further, it has been observed that the nanowire surfaces aid in dislocation multiplication mechanism and
also in dissociation of immobile dislocations.\\

\end{document}